\documentstyle[prl,aps,epsfig,floats]{revtex}

\draft
\begin{document}

\twocolumn[\hsize\textwidth\columnwidth\hsize\csname
@twocolumnfalse\endcsname

\title{Tools for Tunneling}
\author{Uri Sarid\cite{USmail}}
\address{Department of Physics, University of Notre Dame, Notre Dame, IN 46556 USA}
\date{April 16, 1998}
\maketitle

\begin{abstract}
If the universe is trapped and cooled in a metastable false vacuum state, that state will eventually decay by bubble nucleation and expansion. For example, many extensions of the standard model incorporate new scalar fields $\vec\phi$ whose potential has a local minimum at $\vec\phi = 0$ but a global minimum elsewhere, to which the vacuum will eventually tunnel. I calculate the lifetime of the false vacuum, and the field profile of the bubble after tunneling, for any potential that is approximately a polynomial of degree $\le 4$ near the false vacuum. Essentially exact results are given for a single field; for multiple fields a strict lower bound is placed on the tunneling rate.
\end{abstract}
\pacs{PACS numbers: 12.60.Jv, 03.65.Sq, 98.80.Hw
\hfill UND-HEP-97-US01 \hfill hep-ph/9804308}

]

Consider a quantum field theory of some scalar fields $\vec \phi$ (grouped for convenience into a vector) whose potential $U(\vec \phi)$ has local minima at the origin $\vec \phi = 0$ and at some other $\vec \phi = \vec \phi_1$. If $U(\vec \phi_1) < U(0)$ then the vacuum at the origin is a false and metastable one. If the universe is in this false vacuum at some early epoch, and the temperature is sufficiently low -- as I will assume for the remainder of this paper -- then quantum fluctuations will eventually initiate a phase transition from the supercooled initial state to the true vacuum through bubble nucleation and expansion \cite{ref:VolKobOku}. A very elegant and tractable approach to studying this process was introduced by Coleman \cite{ref:ColCalCol}. The bubble's most likely field configuration, called the bounce, is the one which extremizes the Euclidean action, and the lifetime of the false vacuum is proportional to the exponential of this extremal action. The bounce also determines the bubble's form in Minkowski space, after its formation and outside the lightcone of its center.

In this work I calculate the bounce and its Euclidean action for a wide class of models, namely all those for which the potential is at least approximately a polynomial of degree $\le 4$ near the false vacuum. The results are essentially exact for a single field $\phi$; with more fields the action and therefore the false-vacuum lifetime I calculate are strict upper bounds, while the one-dimensional bounce given here may only give a qualitative picture of the true field configuration. One class of applications for this work is to extensions of the standard model containing fields $\vec \phi$ whose vacuum expectation values (VEVs) in the present epoch are phenomenologically required to vanish, for example if they are electrically charged. Thus if a lower minimum of $U$ develops away from the origin, the origin must not only remain a local minimum, but must also be sufficiently long-lived that we would probably still be inhabiting it today. For this type of application the lifetime is of primary importance. It is hoped that, within other contexts, the shape of the bounce will be useful in calculating various other properties of the decay, for example the degree of supercooling, the thickness of the transition region and the distribution of energies within its volume. Of course, for many of these contexts the effects of gravity \cite{ref:GR} and the expansion of the universe, and possibly of finite temperature \cite{ref:FinT}, must be included, but the techniques developed here may readily be extended.

Most previous work on false vacuum decay either assumes a thin-wall approximation to derive analytical results, or specializes to a particular model and numerically extremizes the action. But the former approach is often only applicable when the lifetime is extremely long and hence not very interesting, while the latter can be intricate and time-consuming, requiring a new computation for each model and often also ingenious methods of extremization \cite{ref:Kus}. The results presented here yield either a bound on the action or an essentially exact answer for a very wide class of potentials, with any choice of parameters and without any computer reanalysis.
There is nevertheless some overlap with previous authors. For the case of a single scalar field $\phi$, rescaling was used (as I do below) to exactly study two limiting-case potentials \cite{ref:Lin83,ref:Lin92} and a general quartic potential \cite{ref:ClaHalHin,ref:Shen}, though most of the latter results are given only graphically and for a limited range of parameters; in the high-temperature limit the action was calculated and presented completely in Ref.~\cite{ref:DLHLL}. For the case of multiple fields $\vec \phi$, the reduction to a single field was employed, for example, in Refs. \cite{ref:ClaHalHin,ref:DasDobRan}. These were particular realizations of some of the same techniques used in the present work, and, as further emphasized in Ref.~\cite{ref:Das}, illustrate the usefulness of this approach.

Consider any potential $U$ in any number of scalar fields $\vec \phi$ with a homogeneous false vacuum at $\vec \phi = 0$ and with $U(0) = 0$. In the semiclassical approximation \cite{ref:ColCalCol}, the probability of bubble formation per unit time and volume is $\Gamma/V \simeq m^4 \exp(-S_E)$. The prefactor is dimensionally the fourth power of the typical mass scale in the theory and thus can be readily estimated. I will concentrate exclusively on the argument of the exponential, which is the Euclidean action for the bounce solution. However, care should be taken to account for any large numerical prefactors, for instance \cite{ref:Kus95} group-theoretical factors related to vacuum degeneracy in models with spontaneous symmetry breaking.

The requirement that the false vacuum has probably not decayed in our past light-cone reads: $(\Gamma/V) L^4 \ll 1$, where $L$ is roughly the present size and age of the visible universe. Numerically, $L \sim 10^{10} {\rm yr} \sim 10^{45} {\rm TeV}^{-1}$ so $S_E > 400 + 4 \ln \left(m/{\rm TeV}\right)$. The bounce is an O(4)-symmetric \cite{ref:ColGlaMar} solution of the Euclidean equations of motion which approaches the false-vacuum field configuration $\vec \phi = 0$ at Euclidean time $\tau = \pm \infty$. To calculate it, one must extremize the action with respect to all possible {\it non-trivial} paths through field space which satisfy the boundary conditions. This requires some careful and somewhat time-intensive numerical techniques \cite{ref:Kus}, and the results are difficult to generalize. To simplify this task, consider a single, straight-line path: $\vec \phi = \vec v \phi$ where $\vec v$ is a constant. For example, $\vec v$ can be fixed to point towards the true vacuum, or towards any other direction where tunneling is possible. (The action can also later be minimized with respect to $\vec v$, or to small perturbations thereon \cite{ref:DLHLL}). In any case, the extremum of the true action $S_E$ must be no larger than the extremum of the action $S_E^*$ for $\vec \phi$ restricted to lie along $\vec v$, by a theorem of Coleman \cite{ref:Sid}. Any model not obeying 
\begin{equation}
S_E^* > 400 + 4 \ln \left(m/{\rm TeV}\right)\,.
\label{eq:SEbound}
\end{equation}
will have an even smaller $S_E$ and hence its false vacuum would have decayed by today. Constraining a model based on Eq.~(\ref{eq:SEbound}) isn't as stringent as contraining it based on the full $S_E$, but it is certainly much simpler, can be studied in general as I do below, and is often sufficient to rule out large regions of parameter space. For instance, Dasgupta \cite{ref:Das} has shown for a particular supersymmetric model that most of the parameter range excluded by an exhaustive $S_E$ extremization is already ruled out by the simple $S_E^*$ constraint; see also Ref.~\cite{ref:ClaHalHin}.

Thus we consider a potential function of a single field $\phi$, in particular:
\begin{equation}
U(\phi) = M_2^2 \phi^2 - M_3 \phi^3 + \lambda \phi^4\,.
\label{eq:U}
\end{equation}
This form encompasses all renormalizable potentials, but is also a good approximation for many nonrenormalizable ones which can be expanded as a quartic polynomial near the false vacuum. I assume $M_2^2 > 0$ to ensure at least metastability of the origin, and $M_3 > 0$ without loss of generality. By Euclidean spherical symmetry, the action is
\begin{equation}
S_E = 2 \pi^2 \int_0^\infty \rho^3 \left[ \frac12\,z\,\partial_\mu\phi\partial^\mu\phi + U(\phi) \right]\,d\rho 
\label{eq:SE}
\end{equation}
where $z$ is some constant factor resulting from field rescaling ($z=1$ is canonical normalization) and $\rho = \sqrt{\vec x^2 + \tau^2}$. Using a dimensionless coordinate $x = (M_2/z^{1/2}) \rho$ and field $y(x) = \left(M_3/M_2^2\right)\phi(\rho)$, I find $S_E = \left({z^2 M_2^2/M_3^2}\right) \widehat S_E$, where
\begin{equation} 
\widehat S_E = 2\pi^2 \int_0^\infty x^3 \left[\frac12 \left({dy\over dx}\right)^2 + \widehat U(y)\right]\, dx
\label{eq:SEhat}
\end{equation}
\begin{equation}
\widehat U(y) \equiv y^2 - y^3 + \kappa y^4\,,\quad
\kappa \equiv \lambda M_2^2/M_3^2\,.
\label{eq:Uhat}
\end{equation}
Thus the general problem has been reduced to extremizing an action $\widehat S_E$ depending on a single parameter $\kappa$. Tunneling is possible for any $\frac14 > \kappa > -\infty$; when $\kappa < 0$ some other mechanism must eventually stabilize the potential, but will not influence the following if the true potential is well-approximated by $\widehat U$ at all $0 < y \le y_0$, where $y_0$ is the escape point defined below.

Extremizing $\widehat S_E$ amounts \cite{ref:ColCalCol} to solving the equations of motion
\begin{equation}
{d^2 y\over dx^2}  + {3\over x}{dy\over dx} = - {\partial(-\widehat U)\over \partial y},\quad \left.{dy\over dx}\right|_{x=0} = 0,\quad y(\infty) = 0
\label{eq:eom}
\end{equation}
for a particle located at position $y$ as a function of time $x \ge 0$ and subject to a potential $-\widehat U[y(x)]$ and a time-dependent viscous frictional force. The particle starts at some $y(x = 0) \equiv y_0 > 0$ and slides down the potential $-\widehat U$ monotonically towards the origin, with $y_0$ chosen so $y(x \to \infty) = 0$. 

\begin{figure}
\centerline{\epsfig{figure=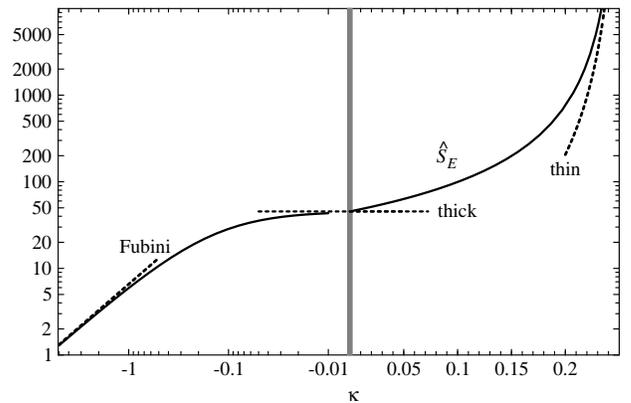,width=3.4in,angle=0}}
\caption{
{\small
The rescaled Euclidean action $\widehat S_E$ as a function of $\kappa$, with asymptotes as dashed lines.}}
\label{fig:SE}  
\end{figure}

\begin{figure}
\centerline{\epsfig{figure=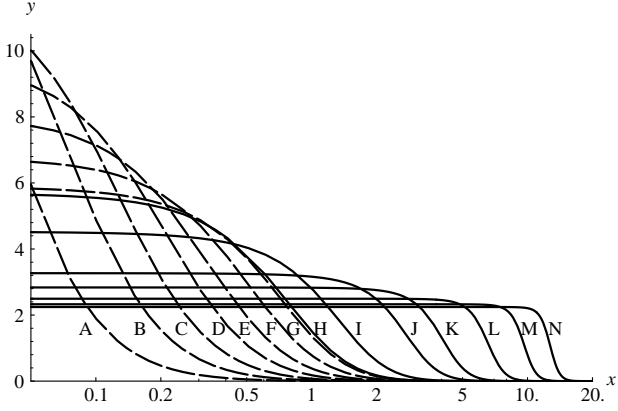,width=3.4in,angle=0}}
\medskip
\caption{
{\small
The (rescaled) bounce $y(x)$ for $0.05 < x < 20$ and various $\kappa$ values: (A) $-5$, (B) $-2$, (C) $-1$, (D) $-0.5$, (E) $-0.25$, (F) $-0.1$, (G) $-0.01$, (H) $0.01$, (I) $0.1$, (J) $0.175$, (K) $0.2$, (L) $0.22$, (M) $0.23$, (N) $0.235$.}}
\label{fig:bounces}  
\end{figure}

\begin{figure}
\centerline{\epsfig{figure=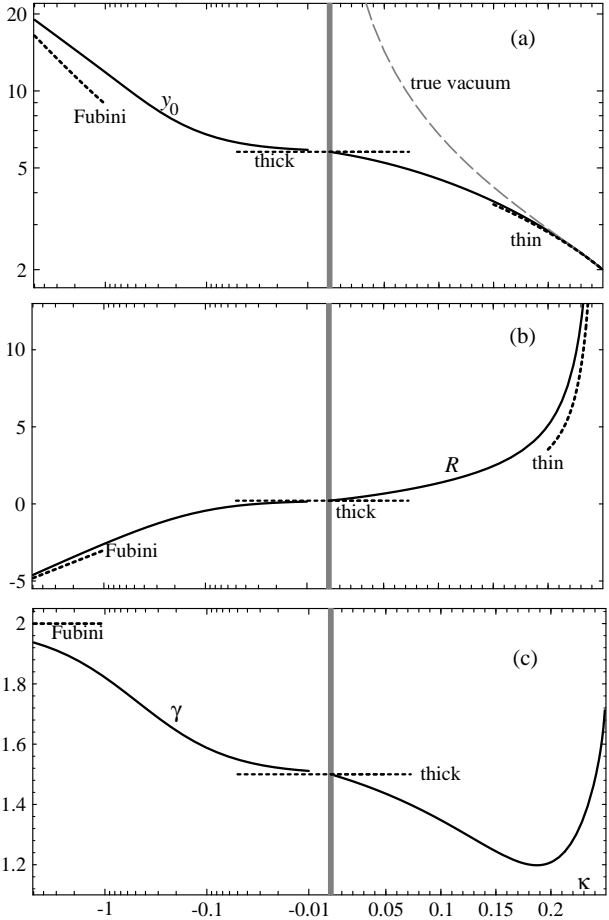,width=3.4in,angle=0}}
\caption{
{\small
The (rescaled) bounce is well-approximated by $y_{\rm fit}(x) = y_0/[1 + x^\gamma e^{\sqrt2 (x - R)}]$; the ``initial'' value $y_0$ and the fit parameters $R$ and $\gamma$ are plotted in (a), (b), and (c), respectively.
}}
\label{fig:fits}  
\end{figure}

The above equations are numerically integrated throughout the allowed range for $\kappa$. The resulting action, from Eq.~(\ref{eq:SEhat}) or equivalently from $\widehat S_E = - 2\pi^2 \int_0^\infty x^3 \widehat U[y(x)]\, dx$, is shown in Fig.~\ref{fig:SE}. To better than 1\% accuracy this action is fit by the semiempirical expressions
\begin{eqnarray}
\widehat S_{E,+} &\simeq& \widehat S_{E,{\rm thick}} - 46.1 +
\widehat S_{E,{\rm thin}} + {16.5\over (1 - 4\kappa)^2} + {28\over 1 - 4\kappa}
\label{eq:SEfitP}\\
\widehat S_{E,-} &\simeq& \widehat S_{E,{\rm thick}} \left[1 + (\widehat S_{E,{\rm thick}}/\widehat S_{E,F})^{1.1} \left|\kappa\right|^{1.1}\right]^{-1/1.1}
\label{eq:SEfitN}
\end{eqnarray}
where the $+$ ($-$) subscript indicates $\kappa > 0$ ($\kappa < 0$), $\widehat S_{E,{\rm thick}} \simeq 45.4$ (see also \cite{ref:Lin83}), $\widehat S_{E,{\rm thin}} = 2\pi^2/[12 (1 - 4 \kappa)^3]$ and $\widehat S_{E,F} = 2 \pi^2/3$ ensure the correct action in the thick wall, thin wall and Fubini asymptotic limits, respectively, as discussed below.

The bounce solutions to the equations of motion are plotted in Fig.~\ref{fig:bounces} for a various $\kappa$. Perhaps surprisingly, they all fit quite accurately the simple form
\begin{equation}
y(x) \simeq y_{\rm fit}(x) \equiv y_0 / \left[1 + x^\gamma e^{\sqrt{2} (x - R)}\right]
\label{eq:yapprox}
\end{equation}
where $y_0 = y(x = 0)$, while $\gamma$ and $R$ are empirically chosen as functions of $\kappa$ to optimize the fit. Fig.~\ref{fig:fits} shows $y_0$ and the best-fit parameters $\gamma$ and $R$ as functions of $\kappa$. Note that when $\kappa$ is positive but significantly below its maximal value, say $\alt 0.15$, the shape of the bubble and especially its action are very different from the thin-wall results. In particular, the field tunnels out to a value $y_0$ far from the true vacuum value $\frac38 \kappa^{-1} \left[1 + (1 - \frac{32}{9} \kappa)^{1/2}\right]$ shown as the long-dashed gray curve in Fig.~\ref{fig:fits}a. The fit parameters are themselves well-approximated by
\begin{eqnarray}
y_{0,+} &\simeq& y_{0,\rm thin} + 1.18 (1 - 4 \kappa)^2 - 1.40 (1 - 4 \kappa)^3
\label{eq:yzerofitP}\\
y_{0,-} &\simeq& y_{0,\rm thick} + 10.4 \ln(1 + |\kappa|) - 1.67 \ln(1 + \kappa^2)
\label{eq:yzerofitN}\\
\gamma_+ &\simeq& 2 - [2 - 7.2 \kappa - 11.4 \kappa^2 - 0.72 \ln(1 - 4 \kappa)]^{-1}
\label{eq:gammafitP}\\
\gamma_- &\simeq& \gamma_F - 1/[2 + 4.9 |\kappa| - 1.3 |\kappa|^{1.3}]
\label{eq:gammafitN}\\
R_+ &\simeq& R_{\rm thin} + 0.91/(1 - 4 \kappa)^{1/2} - 1.41 + 6.4 \kappa^{1.2}
\label{eq:RfitP}\\
R_- &\simeq& 0.21 - 1.13 \ln(1 + 11 |\kappa|^{1.15})
\label{eq:RfitN}
\end{eqnarray}
Here $y_{0,\rm thin} = 2 + 4 (1 - 4\kappa)$, $y_{0,\rm thick} \simeq 5.78$, $\gamma_F = 2$, and $R_{\rm thin} = [\sqrt{2} (1 - 4 \kappa)]^{-1}$ are the asymptotic values described below. The $\sqrt{2} x$ in the exponential makes $y_{\rm fit}$ tend towards zero at the correct rate as $x \to \infty$, and the various expressions (\ref{eq:yzerofitP})-(\ref{eq:RfitN}) give an excellent fit wherever $y(x)$ is significantly different from zero: $|y_{\rm fit}(x) - y(x)|/y_0 \alt 1\%$ for all $x$.

Several limiting cases are of interest. The best-known is the thin-wall regime \cite{ref:ColCalCol}, when $\kappa \to 1/4$ and the two vacua approach degeneracy. To leading order in $1 - 4\kappa$ the equations of motion may be solved analytically: $y_{\rm thin} = y_{0,\rm thin}/\{1 + \exp[\sqrt{2} (x - R_{\rm thin})]\}$. Note that, while in general the solution is better fit by Eq.~(\ref{eq:yapprox}) with finite $\gamma$, in the thin-wall limit $\gamma$ becomes entirely immaterial and $y_{\rm fit} \to y_{\rm thin}$. The opposite limit, in a sense, is when $\kappa \to 0$, which I call (as did Ref.~\cite{ref:KusLanSeg}) the thick-wall regime. The values $\gamma_{\rm thick} \simeq 1.5$ and $R_{\rm thick} \simeq 0.21$ provide a better fit over the entire range of $x$ than the exact solution  of Ref.~\cite{ref:KusLanSeg} to the linearized equations of motion. A more subtle limit is $\kappa \to -\infty$, discussed in some detail in Refs.~\cite{ref:Lin83,ref:Lin92}. Then the cubic term in $\widehat U$ becomes irrelevant, so $\widehat S_E \to |\kappa|^{-1} \widehat S_{E,F}$ where $\widehat S_{E,F} \equiv 2 \pi^2 \int_0^\infty x^3 [\frac12 (dy_F/dx)^2 + y_F^2 - y_F^4] dx$ and $y_F \equiv |\kappa|^{1/2} y$. This action can only be extremized asymptotically. Consider the action of the family of functions $y_{F,\beta} = \sqrt{2} \beta/(\beta^2 + x^2)$. When $\beta \to 0$, the quadratic terms may be neglected, in which case the $y_{F,\beta}$ are exact solutions (known as Fubini instantons \cite{ref:Fub}) with a $\beta$-independent action $\frac23 \pi^2$. Thus when the complete action $\widehat S_{E,F}$ is computed for $y_{F,\beta}$ with $\beta \to 0$, its variation also tends to 0, so $y_{F,\beta}$ becomes an increasingly acceptable semiclassical tunneling solution with an asymptotic extremal action $\widehat S_{E,F} = \frac23 \pi^2$. Returning to the original problem, I expect $\widehat S_E \to |\kappa|^{-1} (\frac23 \pi^2)$ and $y \to |\kappa|^{-1/2} y_{F,\beta\to 0}$, which is largely what I find. (For exponentially large $|\kappa|$ one expects $R \to -2^{-1/2} \ln |\kappa|$; but the fitting function of Eq.~(\ref{eq:RfitN}) was chosen to obtain a simple and adequate fit, not to generate the correct asymptotic behavior.)

The above results allow an easy calculation of the tunneling action, and also of the bubble's field configuration for all $t \ge 0$ and $|\vec x| \ge t$ by using \cite{ref:ColCalCol} the bounce solution with argument $\rho = \sqrt{\vec x^2 - t^2}$. (Within the lightcone, the field typically oscillates about the true vacuum; for this post-tunneling evolution, the equations of motion must be solved anew with imaginary $\rho$.)

One sample application of these results is \cite{ref:RatSar} in the minimal model of supersymmetric gauge mediation, in which the large Higgs VEV hierarchy can generate a new global minimum in the potential of the scalar superpartners of the tau lepton. The potential in fact involves three fields, two of which are electrically charged and so should not acquire a VEV. Using the above methods, and restricting to a straight path connecting the false and true vacua, a significant range of that model's parameters is ruled out on the grounds that tunneling would be too fast. A small span of parameters remains for which a true multidimensional analysis would be needed to test whether the lifetime would be long enough. Note, however, that no vacuum stability study can establish that a parameter value is definitely allowed: even if it produces a long-lived false vacuum, the  evolution of the early universe might put the initial state of $\vec\phi$ near the true vacuum and the model would be unacceptable. Vacuum stability can only rule out parameter values, and the present analysis does just that. 

Somewhat stronger bounds may result not only from tunneling via other paths through a multidimensional field space, but also from thermal fluctuations in the early, hot universe (for a related discussion, see Ref.~\cite{ref:Sher} and references therein). However, such bounds are in general very model-dependent and, in the latter case, may be subject to more uncertainties (see, e.g., Ref.~\cite{ref:Falk}). 

The results of this work provide  significant, robust and easy to determine bounds on a wide class of models in which the acceptable vacuum which we presumably inhabit can be destabilized by quantum tunneling. It is true that for such models the shape of the bounce, also determined in this work, is of only academic interest to us, since we will not survive the tunneling. The analysis of this shape can be useful, however, in a different class of models: ones in which the true vacuum is the one we live in, while the phase transition occurred long ago. Then the profile of the bubble determines various properties of the early universe, such as how and where the latent heat was deposited, what was the spectrum of the particles produced, and what remnants were left behind. To properly study such a scenario may require the inclusion of gravity, the time-dependent evolution of the universe, and possibly nonzero temperature. The results of such studies will quantitatively be quite different from the present work, but hopefully much of the methodology introduced here will still be useful.

Enlightening discussions with S. Coleman, I. Dasgupta, S. Kusenko, A. Linde and especially R. Rattazzi, who collaborated in an early stage of this work, are gratefully acknowledged.


\begin{thebibliography}{99}

\bibitem[*]{USmail}Electronic address: sarid@particle.phys.nd.edu

\bibitem{ref:VolKobOku} M.B. Voloshin, I.B. Kobzarev and L.B. Okun, Sov. J. Nucl. Phys. {\bf 20} (1974) 644.

\bibitem{ref:ColCalCol} S. Coleman, Phys. Rev. D {\bf 15} (1977) 2929; C. Callan and S. Coleman, Phys. Rev. D {\bf 16} (1977) 1762.

\bibitem{ref:GR} S. Coleman and F. De Luccia, Phys. Rev. D {\bf 21} (1980) 3305; S. Parke, Phys. Lett. {\bf B121} (1983) 313.

\bibitem{ref:FinT} A.D. Linde, Phys. Lett. {\bf B100} (1981) 37 and Ref.~\cite{ref:Lin83}.

\bibitem{ref:Kus} A. Kusenko, Phys. Lett. {\bf B358} (1995) 51. See also Ref.~\cite{ref:ClaHalHin} below.

\bibitem{ref:Lin83} A.D. Linde, Nucl. Phys. B {\bf 216} (1983) 421, and {\it Particle Physics and Inflationary Cosmology} (Harwood, Chur, Switzerland, 1990).

\bibitem{ref:Lin92} A.D. Linde, Nucl. Phys. B {\bf 372} (1992) 421.

\bibitem{ref:ClaHalHin} M. Claudson, L.J. Hall and I. Hinchliffe, Nucl. Phys. B {\bf 228} (1983) 501.

\bibitem{ref:Shen} T.C. Shen, Phys. Rev. D {\bf 37} (1988) 3537; K. Enqvist, J. Ignatius, K. Kajantie and K. Rummukainen, Phys. Rev. D {\bf 45} (1992) 3415.

\bibitem{ref:DLHLL} M. Dine, R.G. Leigh, P. Huet, A. Linde and D. Linde, Phys. Rev. D {\bf 46} (1992) 550.

\bibitem{ref:DasDobRan} I. Dasgupta, B.A. Dobrescu and L. Randall, Nucl. Phys. B {\bf 483} (1997) 95; I. Dasgupta, R. Rademacher and P. Suranyi, hep-ph/9804229.

\bibitem{ref:Das} I. Dasgupta, Phys. Lett. {\bf B394} (1997) 116.

\bibitem{ref:Kus95} A. Kusenko, Phys. Lett. {\bf B358} (1995) 47.

\bibitem{ref:ColGlaMar} S. Coleman, V. Glaser and A. Martin, Commun. Math. Phys. {\bf 58} (1978) 211.

\bibitem{ref:Sid} S. Coleman, private communication, based on the correspondence between the bounce and the minimal barrier-penetration path as established in S. Coleman, Nucl. Phys. B {\bf 298} (1988) 178 and Ref.~\cite{ref:ColCalCol}. See also Ref.~\cite{ref:Das}.

\bibitem{ref:KusLanSeg} A. Kusenko, P. Langacker and G. Segre, Phys. Rev. D {\bf 54} (1996) 5824.

\bibitem{ref:Fub} S. Fubini, Nuovo Cimento {\bf 34A} (1976) 521.

\bibitem{ref:RatSar} R. Rattazzi and U. Sarid, Nucl. Phys. B {\bf 501} (1997) 297.

\bibitem{ref:Sher} M. Sher, Phys. Lett. {\bf B317} (1993) 159; Addendum, {\it ibid.} {\bf B331} (1994) 448.

\bibitem{ref:Falk} T. Falk {\it et al.}, Phys. Lett. {\bf B396} (1997) 50.

\end{thebibliography}
\end{document}